\documentclass[onecolumn,11pt]{article}
\usepackage[top=1in, bottom=1in, left=1in, right=1in]{geometry}

\usepackage{times}
\usepackage{sidecap}

\usepackage{dcolumn}
\usepackage{bm}
\usepackage{amsmath}
\usepackage{amssymb}
\usepackage{revsymb}
\usepackage{graphics}
\usepackage{graphicx}
\usepackage{setspace}
\usepackage{color}
\usepackage{wrapfig}

\usepackage{float}
\usepackage{enumitem}

\usepackage[font=small]{caption}

\usepackage{fancyhdr}
\begin{document}

\newpage
\centerline{\textbf{\LARGE{Blackbody-cavity Ideal Solar Absorbers}}}

\centerline{\textbf{Yanpei Tian$^1$, Xiaojie Liu$^1$, Alok Ghanekar$^2$, Fangqi Chen$^1$, Yi Zheng$^{1,3*}$}}

\centerline{$^1$Department of Mechanical and Industrial Engineering, Northeastern University, Boston, MA,USA.}

\centerline{$^2$Artech LLC, Morristown, NJ 07960, USA.}

\centerline{$^3$Department of Electrical and Computer Engineering, Northeastern University, Boston, MA,USA.}
\centerline{$^*$e-mail: y.zheng@northeastern.edu}

\vspace{-10pt}
\section*{\large{Abstract}}
\vspace{-8pt}Spectrally selective solar absorbers (SSAs), harvesting sunlight into heat, are the key to the concentrated solar thermal systems. Current SSAs' designs using photonic crystals, metamaterials, or cermets are either cost-inefficient or have limited applicability due to complicated nanofabrication methods and poor thermal stability at high temperatures. We present a scalable-manufactured blackbody cavity solar absorber design with nearly ideal properties. The unity solar absorptivity and nearly zero infrared emissivity allow for a stagnation temperature of 880$^\circ$C under 10 suns. The performance surpasses those state-of-the-art SSAs manufactured by nanofabrication methods. This design relies on traditional fabricating methods, such as machining, casting, and polishing. This makes it easy for large-scale industrial applications, and the ``blackbody cavity" feature enables its fast-integration to existing concentrated solar thermal systems.

\newpage
\noindent
Solar thermal technology is one of the most promising renewable energy technologies to replace fossil fuel as solar energy is the most abundant form of energy available on earth \cite{boffey1970energy}. Solar absorbers, converting solar radiation into heat are a key component to affect the performance of various solar thermal systems, such as solar thermal power plants and solar thermoelectric generators as well as solar thermophotovoltaics. Ideal SSAs possess a unity solar absorptivity to maximize solar heat gain, while a nearly zero infrared emissivity to minimize energy loss from spontaneous thermal radiation. Photonic crystals \cite{chou2014enabling,chou2014design,yeng2014global}, metamaterials \cite{wang2015highly,wang2012metamaterial,li2018efficient}, and cermet \cite{cao2015high,cao2015enhanced,wang2016polychromic} based spectrally selective solar absorbers have been extensively investigated in the past. However, these approaches often require stringent, complicated, and time-consuming nanofabrication methods such as photolithography, chemical, or physical vapor deposition. Furthermore, cost-effective scaling up of such nanostructures is another hurdle in meeting large scale requirements of possible industrial applications. Even if a simple and low-cost solution-based method is possible for large-scale fabrication of SSAs, the challenge of preparing a uniform and stable precursor containing as-designed metal ions still exists \cite{cao2014review}. Therefore, simple, efficient, stable, and scalable approaches are desirable for commercial solar thermal systems. The feasible material candidates for SSAs are also confined to metals or ceramic \cite{cao2015high,cao2017high,wang2012metamaterial,chou2014enabling,wang2015highly}, such as Ni, W, Cr, Al$_2$O$_3$, SiO$_2$, and Cr$_2$O$_3$, that few fabrication technologies can handle. This affects the large-scale productions and vice versa. Meanwhile, although abovementioned methods achieve good spectral selectivity, today's state-of-the-art SSAs only have selective properties with 90\% $<$ $\alpha_{solar}$ $<$ 98\% and 3\% $<$ $\epsilon_{IR}$ $<$ 10\% \cite{konttinen2003mechanically}. Even an object with 3\% $\epsilon_{IR}$ has a thermal radiation power density of $\sim$ 600 W/m$^2$ at 500$^\circ$C, which causes high thermal loss and reduces the system efficiency. Furthermore, long-term thermal durability under a high-temperature operational environment is critical for engineering applications. The mismatch in the thermal expansion coefficient of metal and ceramic will cause SSAs' fatigue and delamination in accumulated high-temperature thermal cycles \cite{cuomo1975new}. Single-element-based metallic or widely-used metal alloy based SSAs could be a feasible candidate to address this challenge. Therefore, it is highly demanded to fabricate such SSAs with long-term durability through simple traditional fabrication procedures, for instance, machining, casting, soldering, and polishing.

\begin{figure}[!ht]
\centering
\includegraphics[width=1.0\textwidth]{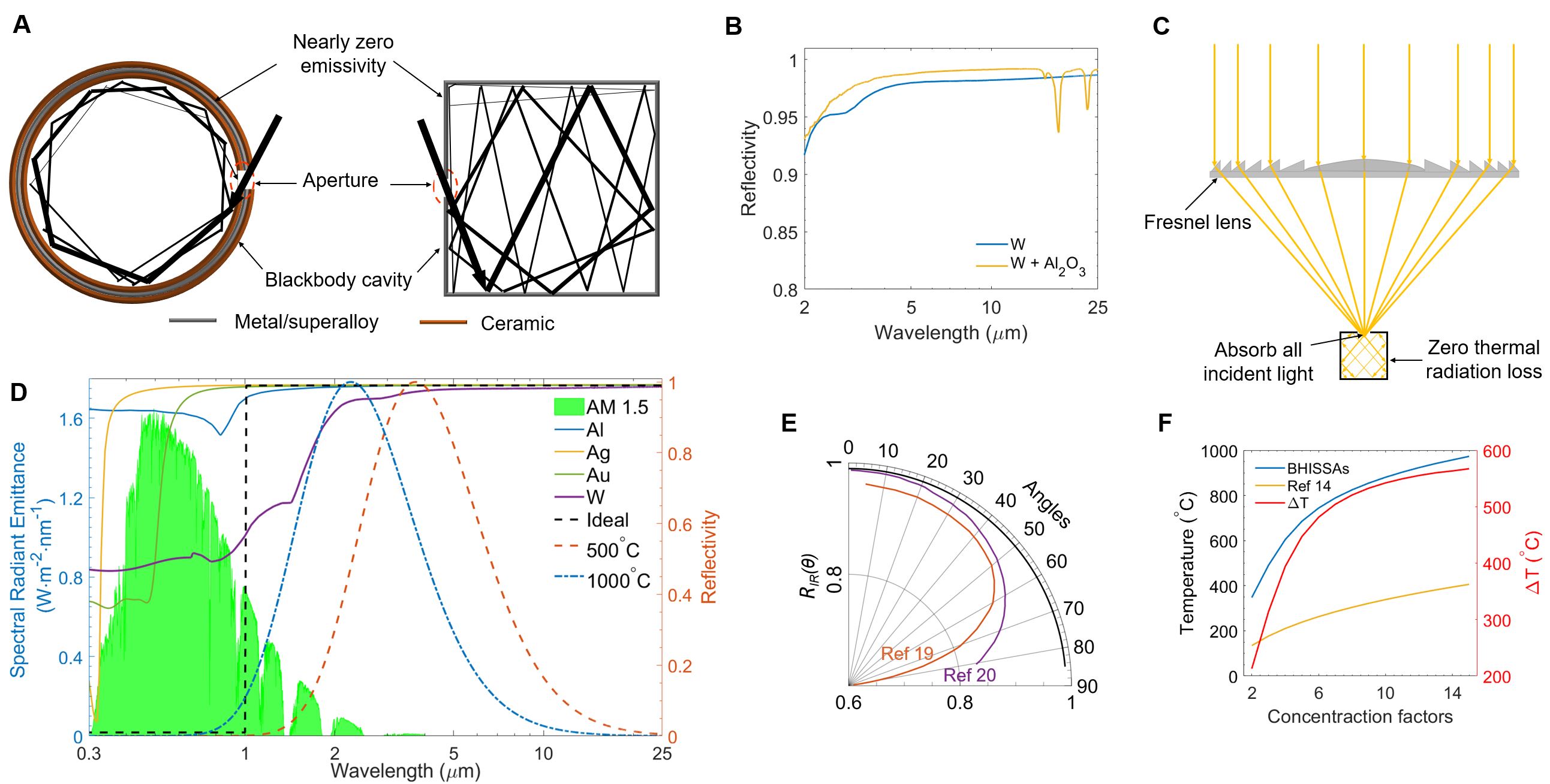}
\caption{\label{fig:Mainfigure} \textbf{Absorbing blackbody cavity with ideal spectral selectivity.} (\textbf{A}) Left, ceramic sandwiched BISAs to be anti-oxidation at high operational temperatures. Right, all-metal or superalloy based BISAs. (\textbf{B}) Spectral reflectivity of the W without and with a 100 nm thick ceramic (e.g. Al$_2$O$_3$) coating.(\textbf{C}) Schematic of a blackbody cavity light absorbing system showing the cubic BISAs integrated with the existing concentrated solar thermal systems using Fresnel lens. (\textbf{D}) Spectral reflectivity (1 - $\epsilon$) of the Al, Ag, Au, W and ideal SSAs displayed with the ASTM G173 solar spectrum and the normalized blackbody irradiance at 500 $^\circ$C and 1,000 $^\circ$C, respectively. (\textbf{E}) Nearly unity reflectivity $R_{IR}$ across angles of BISAs's external surface when fabricated by metal (Here, W is used to calculate and average it from 1 $\mu$m to 25 $\mu$m). (\textbf{F}) Left $y$ axis: Stagnation temperature as a function of concentration factors for BISAs and the cermet based SSAs fabricated by Cao et al \cite{cao2017high} (the convection heat transfer coefficient, $h$, is 5 W/m$^2$ K). Right $y$ axis: The difference of stagnation temperatures between BISAs and the cermet based SSAs in reference 14.
} 
\end{figure}

When a spherical chamber at thermal equilibrium with a small aperture, it can be regarded as an ideal blackbody model \cite{hagen1999heat}, since the incident light undergoes reflection and absorption of many times, and the light intensity will decrease partly each time when it absorbs by the internal surface. Eventually, light scarcely escapes from the spherical chamber. Therefore, the small cavity shares the same properties with a blackbody. The effective absorptivity of the blackbody model gets even bigger at a fixed area ratio ($\phi$) of the blackbody cavity and the chamber surface area when the absorptivity of the chamber material increases. If $\phi$ $<$ 0.6\% and the internal surface absorptivity is 0.6, the overall absorptivity of the blackbody cavity is 0.996. In principle, the absorptivity of materials for the chamber does not affect the final absorptivity of the ``blackbody cavity" (Fig. \ref{fig:Mainfigure} \textbf{A}, left), that is, even if the material of chamber is high reflective metal, like Al, Ag, and W, etc, the overall absorptivity of the blackbody cavity can still approach to be unity with an optimal area ratio. The absorptivity of the blackbody cavity is also independent of incident angles. The shape of the blackbody chamber can also be cubic (Fig. \ref{fig:Mainfigure} \textbf{A}, right), cylindrical, and conical as well as any shapes according to the demand for practical applications. The blackbody-cavity ideal solar absorbers (BISAs) can be fabricated using metal directly since the metals have nearly unity reflectivity in the infrared range which yields that the outer surface of the blackbody chamber sacredly has thermal loss due to spontaneous thermal radiation. The metals can also be sandwich-coated with ceramic materials (e.g. SiO$_2$, Al$_2$O$_3$, etc.) or use superalloy (e.g. Inconel, Waspaloy, etc.) to become oxidation-resistant for their use under a high-temperature environment. For example, 100 nm thick Al$_2$O$_3$ coated W shows little change or even an increased reflectivity from 2 $\mu$m to 25 $\mu$m (Fig. \ref{fig:Mainfigure} \textbf{B}, calculation approaches of reflectivity spectra are provided in the supplementary information \cite{chew1995waves,narayanaswamy2014infrared,moharam1995stable}). Besides, the concentrating optics used in concentrated solar thermal systems converge the incident sunlight to a small point, and hence it makes the blackbody cavity perfect to be integrated to the existing solar thermal systems using Fresnel lens or parabolic reflector as concentrating optics without complex modifications (Fig. \ref{fig:Mainfigure} \textbf{C}). Figure. \ref{fig:Mainfigure} \textbf{D} shows that some common metals, like Al, Ag, Au, Ni, and W, have nearly unity reflectivity in the thermal infrared region (2 $\mu$m to 25 $\mu$m). Consequently, these metals or related alloys can be selected as alternative materials for different temperature applications according to their melting and oxidation temperatures. Furthermore, these metals like W shows an excellent reflectivity across the incident angle (0$^\circ$ $\sim$  85$^\circ$) that is higher than previously reported values (Fig. \ref{fig:Mainfigure} \textbf{E}) \cite{tesfamichael2000angular,zhu2009optical}. The novel design of BISAs results in approaching ideal spectral selectivity, exemplified by a stagnation temperature of 880$^\circ$C under a concentration factor of 10 (10 $\times$ AM 1.5 solar intensity). It is 542$^\circ$C higher than the cermet based SSAs' stagnation temperature (the reflectivity spectra used in the calculation of stagnation temperature are taken from reference 13, Fig. \ref{fig:Mainfigure} \textbf{F}, further details about the stagnation temperature calculation are provided in supplementary information).

Above all, we demonstrate a nearly perfect SSAs with unity solar absorptivity and nearly zero infrared thermal emissivity by employing a novel blackbody cavity thermal radiation model. Both the solar absorptivity and thermal emissivity of the blackbody cavity is angle-independent. Additionally, we demonstrate a stagnation temperature of 880$^\circ$C under 10 suns. The BISAs can be fabricated by machining, casting, or soldering into different shapes with metals or superalloys and then polishing its external surface into a highly reflective ``mirror" so as to suppress the energy loss from spontaneous thermal radiation. This novel design of an absorbing blackbody cavity with a scalable-manufactured feature makes it perfect to be simply integrated into existing concentrated thermal systems.

\section*{Data availability}
The datasets analyzed during the current study are available
within the paper. All other data related to this work are available from the corresponding author upon reasonable request.

\newpage
\bibliography{Yanpei}

\newpage
\section*{Acknowledgments}
This project is supported by the National Science Foundation through grant number CBET-1941743.

\section*{Author contributions}
Y.T., X.L., and Y.Z. develop this concept. Y.T. develops the theoretical model and writes the manuscript with help from all other authors. X.L., A.G., and F.C. help revised the manuscript. Y.Z. supervises this project.

\section*{Competing interests}
The authors declare no conflict of interest.

\end{document}